\documentclass[superscriptaddress,nobibnotes,prb,twocolumn]{revtex4-2}
\usepackage{graphicx}
\usepackage{bm}
\usepackage{color}
\usepackage[squaren]{SIunits}
\usepackage{amssymb}
\usepackage{epstopdf}
\usepackage{natbib}
\usepackage[colorlinks,citecolor=blue]{hyperref}
\usepackage{amsmath}

\begin{document}
	
	\title{Two-dimensional superconductivity in new niobium dichalcogenides-based bulk superlattices}
	\author{Kaibao Fan}
	\affiliation{Hefei National Research Center for Physical Sciences at the Microscale, University of Science and Technology of China, Hefei, Anhui 230026, China}
	
	\author{Mengzhu Shi}
	\altaffiliation{E-mail: mzshi@ustc.edu.cn}
	\affiliation{Hefei National Research Center for Physical Sciences at the Microscale, University of Science and Technology of China, Hefei, Anhui 230026, China}
	
	\author{Houpu Li}
	\affiliation{Department of Physics and CAS Key Laboratory of Strongly-coupled Quantum Matter Physics, University of Science and Technology of China, Hefei, Anhui 230026, China}
	
	\author{Ziji Xiang}
	\affiliation{Hefei National Research Center for Physical Sciences at the Microscale, University of Science and Technology of China, Hefei, Anhui 230026, China}
	\affiliation{Hefei National Laboratory, University of Science and Technology of China, Hefei, Anhui 230026, China}
	
	\author{Xianhui Chen}
	\altaffiliation{E-mail: chenxh@ustc.edu.cn}
	\affiliation{Hefei National Research Center for Physical Sciences at the Microscale, University of Science and Technology of China, Hefei, Anhui 230026, China}
	\affiliation{Department of Physics and CAS Key Laboratory of Strongly-coupled Quantum Matter Physics, University of Science and Technology of China, Hefei, Anhui 230026, China}
	\affiliation{Hefei National Laboratory, University of Science and Technology of China, Hefei, Anhui 230026, China}
	
	\begin{abstract}
		Transition metal dichalcogenides exhibit many unexpected properties including two-dimensional (2D) superconductivity as the interlayer coupling being weakened upon either layer-number reduction or chemical intercalation. Here we report the realization of 2D superconductivity in the newly-synthesized niobium dichalcogenides-based bulk superlattices Ba$_{0.75}$ClNbS$_{2}$ and Ba$_{0.75}$ClNbSe$_{2}$, which consists of the alternating stacking of monolayer $H$-NbS$_{2}$ (or $H$-NbSe$_{2}$) and monolayer inorganic insulator spacer Ba$_{0.75}$Cl. Magnetic susceptibility and resistivity measurements show that both superlattices belong to type-II superconductor with $T_{c}$ of 1 K and 1.25 K, respectively. Intrinsic 2D superconductivity is confirmed for both compounds below a Berezinskii-Kosterlitz-Thouless transition and a large anisotropy of the upper critical field. Furthermore, the upper critical field along $ab$ plane ($H_{c2}^{\parallel ab}$) exceeds the Pauli limit ($\mu_{0}H_{p}$) in Ba$_{0.75}$ClNbSe$_{2}$, highlighting the influence of spin-orbit interactions. Our results establish a generic method for realizing the 2D superconducting properties in bulk superlattice materials.		
		
\end{abstract}
	
	\maketitle

	\section{Introduction}
	
	Two-dimensional (2D) superconductivity has been widely concerned and its study has provided more insights into various quantum phenomena. An array of novel quantum states in low-dimensional physics of atomically thin superconductors have been continuously explored and intensively discussed through rich theoretical and experimental work \cite{ref1,ref2,ref3,ref4,ref5,ref6}. A multitude of exotic quantum phenomena such as quantum Griffiths singularity \cite{ref7}, anomalous quantum metallic state \cite{ref8}, the Berezinskii-Kosterlitz-Thouless (BKT) transition \cite{ref9,ref10} and Ising superconductivity \cite{ref11} arise as superconducting materials approach the 2D limit. Highly crystalline 2D superconductors including interfacial superconductors \cite{ref12,ref13}, ion-gated superconductors \cite{ref14,ref15}, highly crystalline atomic layers grown by molecular beam epitaxy \cite{ref16} and mechanically exfoliated 2D crystals \cite{ref17,ref18} provide an ideal platform for searching the quantum phase transitions in low dimension. Apart from the above mentioned thin flake or thin film samples, 2D superconductivity can also emerge in bulk crystals. The recently reported Ba$_{6}$Nb$_{11}$S$_{28}$ \cite{ref19} with thick insulating block between superconducting layers and SrTa$_{2}$S$_{5}$ \cite{ref20} with unidirectional structure modulation are exemplary  bulk superlattices hosting 2D superconductivity and the corresponding novel phenomena due to dimension reduction --- Fulde-Ferrell-Larkin-Ovchinnikov (FFLO) state, pair density wave, etc. This suggests that enriched 2D characteristics can be integrated into layered superconductors by modulating the interlayer interaction.

    At present, artificial design is an effective route to extend the superconducting material system, the most celebrated examples including the twisted graphene \cite{ref21,ref22}, superconducting heterojunction \cite{ref4,ref23}, and the commensurate or incommensurate bulk superlattices \cite{ref24,ref25}, etc. In the rich family of 2D materials, layered transition metal dichalcogenides (TMDs) become the focus of fundamental research and technological applications due to their unique crystal structures \cite{ref26} and tunable physical properties \cite{ref27,ref28}. The TMD materials exhibit distinct behaviors between their atomic layers and bulk crystals owing to the dimension-dependent electronic structure \cite{ref18,ref29}. For instance, a large in-plane upper critical field ($H_{c2}^{\parallel ab}$) far exceeding Pauli limit ($\mu_{0}H_{p}$) has been reported in the exfoliated NbSe$_{2}$ and TaS$_{2}$ \cite{ref18,ref30}. In these monolayer TMD superconductors, the breaking of inversion and (or) mirror symmetries possibly leads to a unique Ising pairing arising from valley-dependent spin-orbit-coupling (SOC) \cite{ref28}, which increases the in-plane upper critical field dramatically. However, compared with the complexity of obtaining and studying monolayer samples, it is particularly important to achieve 2D characteristics directly in bulk crystalline materials. A viable strategy to weaken the interlayer coupling is to introduce the large ions or the insulating layers into van der Waals bulk crystalline materials to extend the adjacent interlayer spacing \cite{ref29,ref31,ref32,ref33}, instead of reducing the sample thickness.

    In this work, we report the structural characterizations and superconducting properties in newly-discovered niobium dichalcogenides-based superlattice Ba$_{0.75}$ClNbS$_{2}$ and Ba$_{0.75}$ClNbSe$_{2}$. The as-synthesized superlattices consist of the alternating stacking of monolayer $H$-NbS$_{2}$ (or $H$-NbSe$_{2}$) and monolayer inorganic insulator spacer Ba$_{0.75}$Cl, which results in a strong decoupling between the adjacent superconducting layers.  Magnetic susceptibility and resistivity measurements confirm the bulk superconductivity in Ba$_{0.75}$ClNbS$_{2}$ and Ba$_{0.75}$ClNbSe$_{2}$ with $T_{c}$ of 1 K and 1.25 K, respectively. The nature of 2D superconductivity is revealed in both compounds, underlined by a BKT transition. Transport measurements show that both superlattices exhibit a large anisotropy of the upper critical field. Furthermore, the in-plane upper critical field $H_{c2}^{\parallel ab}$ exceeds the Pauli limit $\mu_{0}H_{p}$ in Ba$_{0.75}$ClNbSe$_{2}$, emphasizing the effect of spin-orbit interactions of heavy atoms. Our findings provide a valuable platform to achieve the 2D superconductivity and study exotic quantum phases in TMDs-based bulk superlattices.

	\section{Experimental Methods}
	
    The synthesis of the single crystal Ba$_{0.75}$ClNbS$_{2}$ and Ba$_{0.75}$ClNbSe$_{2}$ was achieved by a BaCl$_{2}$-flux method. Firstly, the anhydrous BaCl$_{2}$ powder was dried at 700 ℃ for 20 h before using. Secondly, Ba (99.5\%) flakes, Nb powder (99.9\%), S or Se powder (99.999\%), and anhydrous BaCl$_{2}$ powder (99.9\%) were mixed with the molar ratio of 0.6:1:2:25 in a glove box filled with Argon gas. The mixture was loaded into an alumina crucible and sealed in a double-walled quartz ampoule. The above ampoule was heated to 1150 ℃ in 24 h, kept at 1150 ℃ for 24 h and then slowly cooled down to 800 ℃ in 7 days, followed by a power-off cooling. Single crystals with typical dimensions of 0.4$\times$0.5$\times$0.02 mm$^{3}$ were obtained after soaking the product in dimethyl formamide (99.9\%) for 2 days in a glove box filled with Argon gas.
    
    Crystal structures were solved from the four-circle X-ray diffraction data collected at 300 K with Cu $K_{\alpha}$ radiation (Rigaku XtaLAB PRO 007HF) in Core Facility Center for Life Sciences, USTC; the data were processed and reduced using CrysAlisPro \cite{ref34} and the structures were solved and refined using Olex-2 \cite{ref35} with the ShelXT and ShelXL packages \cite{ref36,ref37}. Further details of the crystal structure could be obtained from the joint CCDC/FIZ Karlsruhe online deposition service with number CSD 2395029 and 2395036. Furthermore, the powder X-ray diffractometer (Rigaku SmartLab 9) with Cu $K_{\alpha}$ radiation was used to select larger crystals for transport measurements. Chemical composition  was  determined  using a Hitachi SU8220 field emission scanning electron microscope (FE-SEM) equipped with an energy-dispersive X-ray spectroscopy (Oxford Instrument X-MaxN 150). The cross-section high angel annular dark-field (HAADF) image was collected using a Talos F200X microscope with a working voltage of 200 kV. 
    
    Longitudinal resistivity and Hall measurements were conducted in a Quantum Design Physical Property Measurement System (PPMS, 9T) equipped with a $^{3}$He system. All of the in-plane electrical transport data were measured in a standard Hall-bar configuration, whereas the out-of-plane resistivity was acquired using four-probe measurements based on a Corbino disk configuration. Magnetic susceptibility was measured in a Quantum Design Magnetic Property Measurement System (MPMS-5) equipped with a $^{3}$He insert; $H$ was applied parallel to the $ab$ plane during measurements.

	\begin{figure*}[htp]
	\centering
	\includegraphics[width=0.9\textwidth]{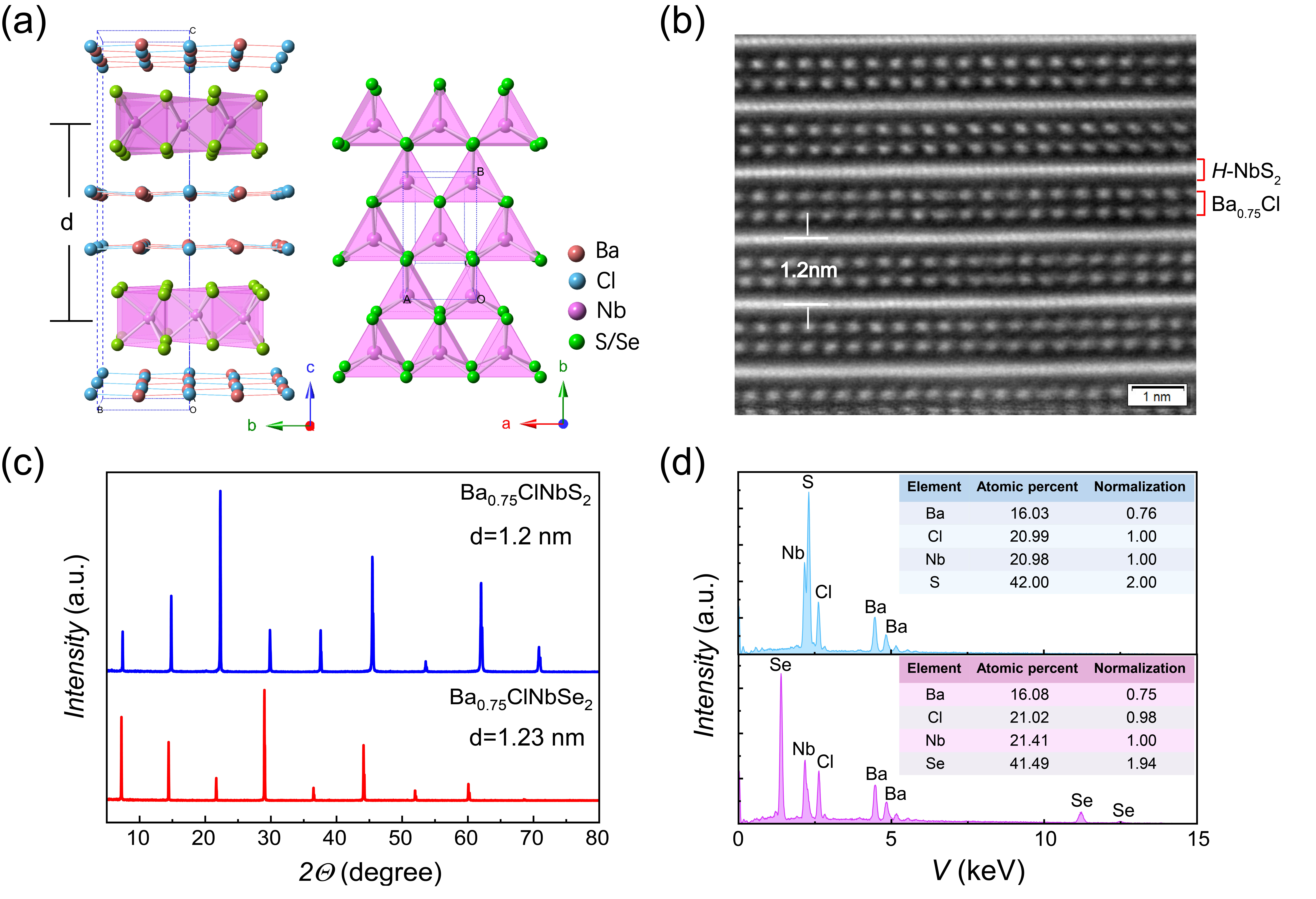}
	\caption{Structure and composition of Ba$_{0.75}$ClNbS$_{2}$ and Ba$_{0.75}$ClNbSe$_{2}$. (a) Crystal structure of Ba$_{0.75}$ClNbS$_{2}$/Se$_{2}$ (left) and in-plane structure of $H$-NbS$_{2}$/NbSe$_{2}$ layer (right). The value of d is the spacing between adjacent TMD layers. Blue dashed lines denote the superlattice unit cell. (b) The cross-section HAADF image of Ba$_{0.75}$ClNbS$_{2}$ single crystal. The scale bar is 1 nm. (c) Single crystal (SC-)XRD patterns for Ba$_{0.75}$ClNbS$_{2}$ and Ba$_{0.75}$ClNbSe$_{2}$; $(00l)$ series of diffraction peaks are clearly observed. (d) The EDS analysis for Ba$_{0.75}$ClNbS$_{2}$ and Ba$_{0.75}$ClNbSe$_{2}$ single crystal. The inset shows chemical composition ratio of the corresponding material. The element ratio Ba : Cl : Nb : S/Se is close to 0.75 : 1 : 1 : 2, with Ba : Cl : Nb : S = (0.76 $\pm$ 0.01) : (1.00 $\pm$ 0.02) : 1 : (2.00 $\pm$ 0.04) and Ba : Cl : Nb : Se = (0.75 $\pm$ 0.02) : (0.98 $\pm$ 0.03) : 1 : (1.94 $\pm$ 0.05).
	}
	\label{FIG. 1}
\end{figure*}

	\section{Results and Discussion}
	
    The crystal structure and composition of the as-synthesized superlattices of Ba$_{0.75}$ClNbS$_{2}$ and Ba$_{0.75}$ClNbSe$_{2}$ were determined through a combination of powder X-ray diffraction (pXRD), four-circle single crystal X-ray diffraction (SC-XRD), high-angle annular dark field (HAADF) imaging, energy dispersive X-ray spectrum (EDS) measurements. Figure. 1(a) shows the crystal structure of Ba$_{0.75}$ClNbS$_{2}$/Se$_{2}$, which can be viewed as an alternate stacking of the Ba-Cl spacer layer and the $H$-NbS$_{2}$/NbSe$_{2}$ layer; the presence of spacing layer strongly suppress the interlayer coupling between TMD sheets. In the $H$-type TMD layers, each Nb atom forms chemical bonds with six adjacent S or Se atoms, forming a triangular-prism coordination polyhedron. The specific structure solved from the SC-XRD is shown in the Table S1 \cite{ref38}. The cross-section HAADF image of Ba$_{0.75}$ClNbS$_{2}$ in Fig. 1(b) displays a clear alternating stack of $H$-NbS$_{2}$ and Ba-Cl sublayers. A series of sharp ($00l$) diffraction peaks [Fig. 1(c)] suggest the high quality of both single crystals. The $c$-axis cell parameters obtained from the ($00l$) diffraction peaks are consistent with the results of SC-XRD. The introduction of Ba-Cl layer expands the spacing of the adjacent TMD layers to 12 Å (12.3 Å) in Ba$_{0.75}$ClNbS$_{2}$ (Ba$_{0.75}$ClNbSe$_{2}$), which is remarkably larger than the parent 2$H$-NbS$_{2}$ and 2$H$-NbSe$_{2}$ (5.71 Å and 6.27 Å, respectively). The two-dimensionality of the superlattice structure is substantial enhanced compared to the parent structure due to a weaker interlayer interaction. The EDS mapping and analysis [Fig. S1 and Fig. 1(d)] confirm the homogeneity of both single crystal samples and the composition ratio of Ba : Cl : Nb : S (Se) close to 0.75 : 1 : 1 : 2.

	\begin{figure*}[htp]
		\centering
		\includegraphics[width=0.9\textwidth]{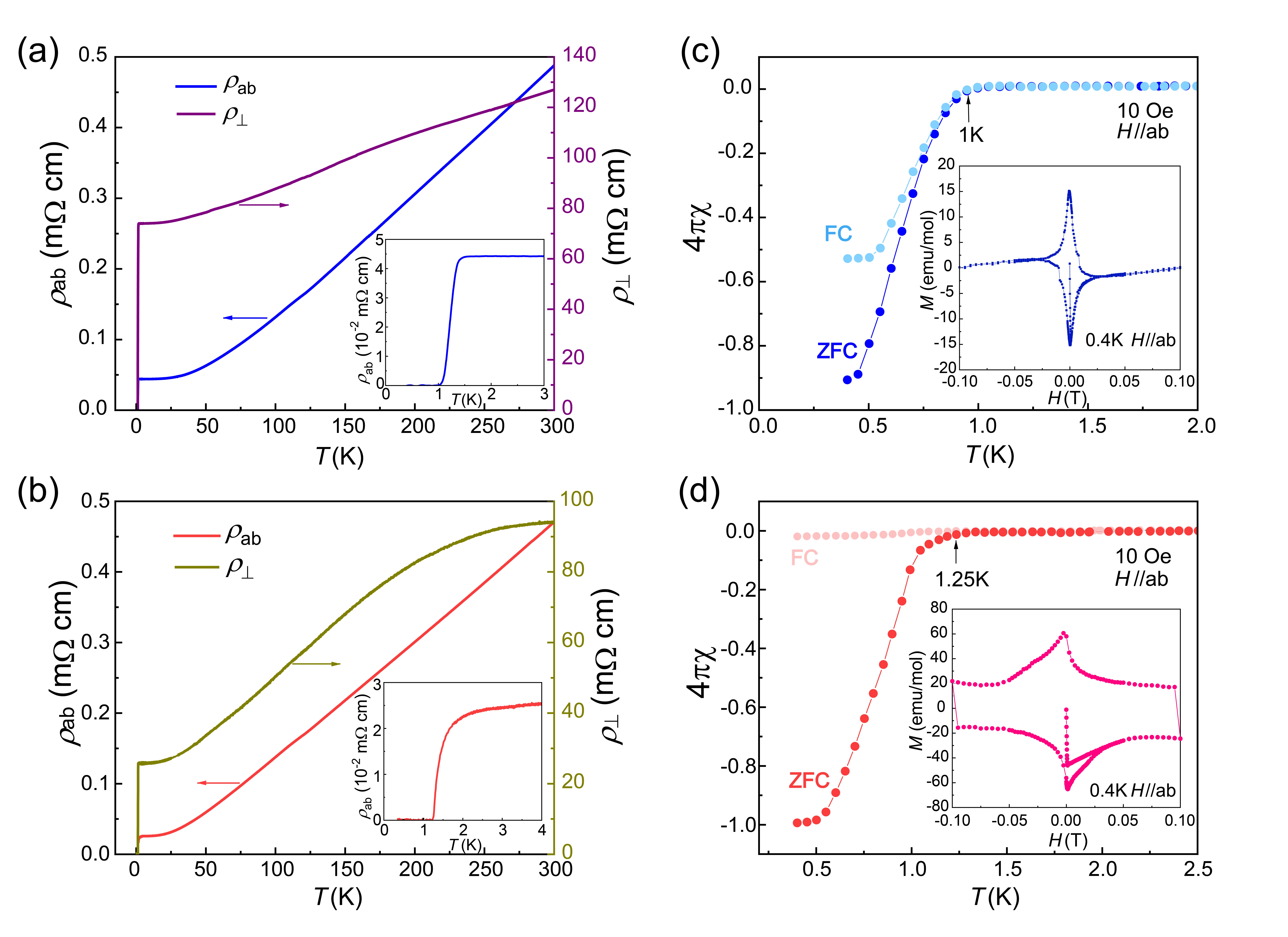}
		\caption{Resistivity and superconducting transition of Ba$_{0.75}$ClNbS$_{2}$ and Ba$_{0.75}$ClNbSe$_{2}$. (a) and (b) Temperature-dependent in-plane resistivity ($\rho_{ab}$) and out-of-plane resistivity ($\rho_{\perp}$) measured in Ba$_{0.75}$ClNbS$_{2}$ (a) and Ba$_{0.75}$ClNbSe$_{2}$ (b), respectively. Insets show expanded views of the superconducting transition. (c) and (d) Magnetic susceptibility curves measured with an in-plane magnetic field of 10 Oe for Ba$_{0.75}$ClNbS$_{2}$ (c) and Ba$_{0.75}$ClNbSe$_{2}$ (d), respectively. Zero-field cooled (ZFC) and field-cooled (FC) curves are displayed in different colors. Insets display the magnetic hysteresis loop in the superconducting state, measured under in-plane magnetic fields at 0.4 K for Ba$_{0.75}$ClNbS$_{2}$ and Ba$_{0.75}$ClNbSe$_{2}$.
}
\label{FIG. 2}
	\end{figure*}

    The transport properties of Ba$_{0.75}$ClNbS$_{2}$ and Ba$_{0.75}$ClNbSe$_{2}$ single crystals with high quality have been systematically characterized. Figure 2(a) and 2(b) show the temperature dependence of the in-plane ($\rho_{ab}$) and out-of-plane ($\rho_{\perp}$) resistivity of Ba$_{0.75}$ClNbS$_{2}$ and Ba$_{0.75}$ClNbSe$_{2}$ at zero field, respectively. For both compounds, $\rho_{ab}$  and $\rho_{\perp}$ exhibit a metallic behavior and tend to saturation below 20 K. Furthermore, the superconducting transition occurs successively at lower temperature. The embedded illustrations show the zero-resistance temperature ($T_{c}$) of Ba$_{0.75}$ClNbS$_{2}$ (Ba$_{0.75}$ClNbSe$_{2}$) is about 1 K (1.25 K), which is consistent with the result of magnetic susceptibility in Fig. 2(c) and 2(d). The nearly 100\% shielding volume fraction hints at the bulk superconductivity of the both materials. The $T_{c}$ of our superlattices is significantly lower than that of the parent 2$H$-NbS$_{2}$ and 2$H$-NbSe$_{2}$ ($T_{c} \sim$ 6 K and 7 K \cite{ref39,ref40}, respectively), aligning with the observed trend in the parent's thin layer wherein the $T_{c}$ diminishes with the reduction in thickness within the parent structure \cite{ref30,ref41}. This implies that it is feasible to attain the characteristics akin to those of few or even single layers in such superlattices.  
    
    An increase in the inter-layer spacing between the conducting layers naturally induces a more pronounced 2D character of the electronic system, resulting in significant anisotropy in electrical transport properties. At low $T$, the resistivity anisotropy $\rho_{\perp}$/$\rho_{ab}$ reaches the order of 10$^{3}$ (Fig. S2) in Ba$_{0.75}$ClNbS$_{2}$ and Ba$_{0.75}$ClNbSe$_{2}$, which is about ten times larger than that of the parent materials (Table S2) \cite{ref42,ref43} corroborating the enhanced 2D nature of the electronic structure in our superlattices. The Hall resistivity at various temperature [Fig. S3(c) and 3(d)] shows a linear behavior with a positive slope, suggesting the hole-type carriers dominated transport behavior. The carrier density $n_{H}$ and Hall mobility $\mu_{H}$ can be deduced from the Hall resistivity $\rho_{xy}: n_{H} = (eR_{H})^{-1}$, $\mu_{H} = (1/\mu_{0}H)(\rho_{xy}/\rho_{xx})$, where $R_{H} = \rho_{xy}/\mu_{0}H$ is the Hall coefficient, $e$ is the electron charge, $\mu_{0}$ is the vacuum permeability. Based on the Hall data at 5 K [Fig. S3(e) and 3(f)], the $n_{H}$ and $\mu_{H}$ can be estimated: $n_{H}$ =2.5 (4.4)$\times$10$^{21}$ cm$^{-3}$, $\mu_{H}$ = 23 (130) cm$^{2}$ V$^{-1}$ s$^{-1}$ for Ba$_{0.75}$ClNbS$_{2}$ (Ba$_{0.75}$ClNbSe$_{2}$).

\begin{figure*}[htp]
	\centering
	\includegraphics[width=0.9\textwidth]{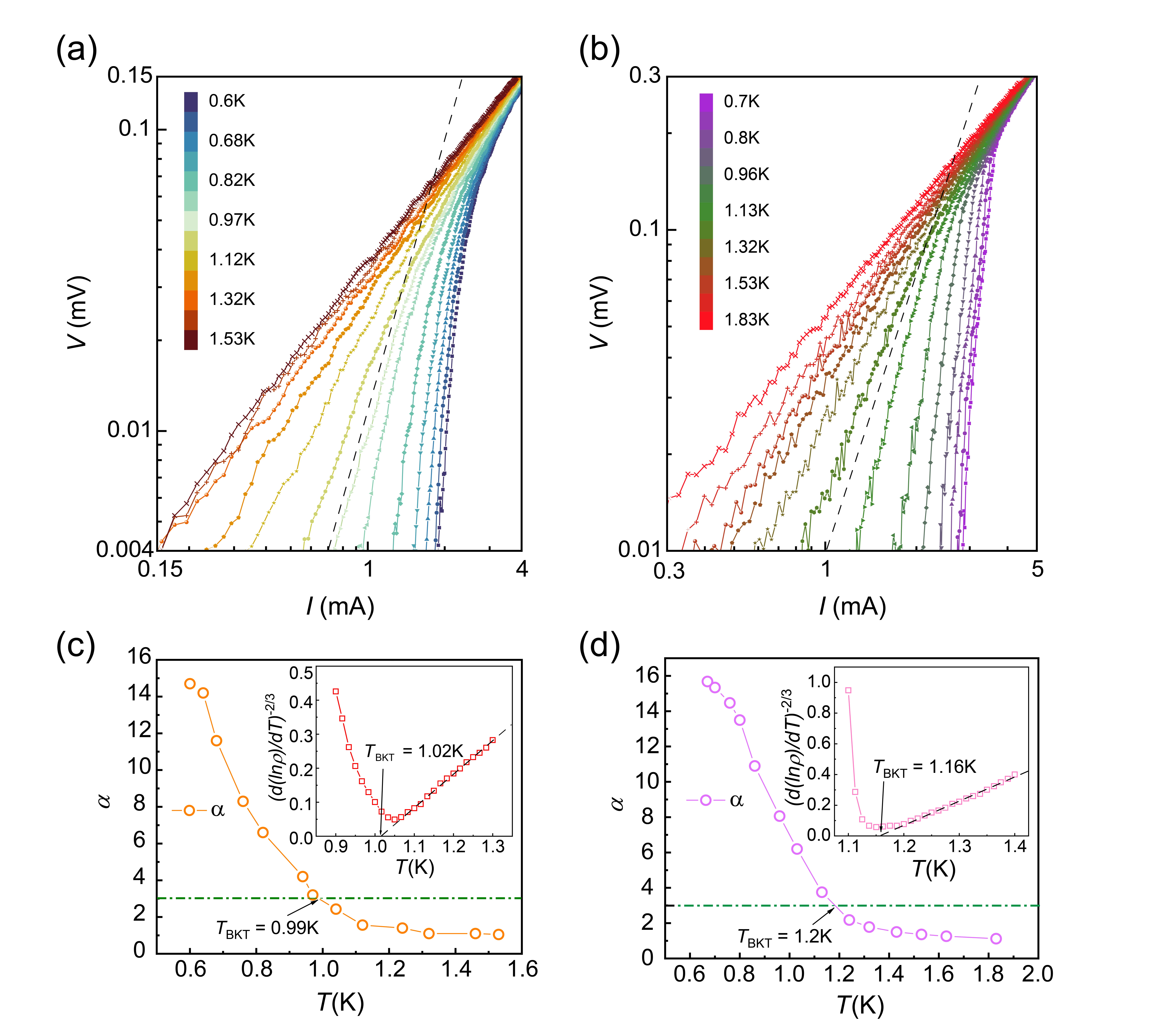}
	\caption{The 2D superconductivity of Ba$_{0.75}$ClNbS$_{2}$ and Ba$_{0.75}$ClNbSe$_{2}$. (a) and (b) Current-voltage characteristics at zero field at various temperatures plotted in a logarithmic-logarithmic scale for Ba$_{0.75}$ClNbS$_{2}$ (a) and Ba$_{0.75}$ClNbSe$_{2}$ (b). The black dash lines indicate the power-law behavior $V \propto I^{3}$. (c) and (d) Temperature dependence of the power-law exponent $\alpha$ deduced from the $V \propto I^{\alpha}$ relationship in Ba$_{0.75}$ClNbS$_{2}$ (c) and Ba$_{0.75}$ClNbSe$_{2}$ (d). The BKT transition temperature $T_{BKT}$ defined as the temperature at which $\alpha$ = 3 (green dash dot lines) is 0.99 K and 1.02 K for Ba$_{0.75}$ClNbS$_{2}$ and Ba$_{0.75}$ClNbSe$_{2}$, respectively. Insets show resistivity curves around $T_{c}$ with the $y$-axis plotted in the $[dln(\rho)/dT]^{-2/3}$ scale for Ba$_{0.75}$ClNbS$_{2}$ and Ba$_{0.75}$ClNbSe$_{2}$, respectively. Dashed lines are linear fits. The BKT transition temperature $T_{BKT}$ is also determined for both compounds.
	}
	\label{FIG. 3}
\end{figure*}

    The remarkably enhanced two-dimensionality in the superlattices not only renders the transport properties highly anisotropic in the normal state, but also results in a remarkable 2D superconductivity. In the 2D limit, any long-range orders with phase coherence (occurring concurrently with a spontaneous breaking of a continuous symmetry) will be destroyed due to the strong fluctuation in the phase field of the order parameter. Nevertheless, BKT phase transition can still happen, at which quasi-long-range order emerges without spontaneous symmetry breaking \cite{ref9,ref44}. Hence in 2D superconductors, the superconducting transition can be described as a BKT transition from unpaired vortices and antivortices to bound vortex–antivortex pairs \cite{ref45,ref46}. Below the transition temperature $T_{BKT}$, the current-voltage ($I$-$V$) relationship follows a power-law behavior given by $V \propto I^{\alpha}$ with the exponent $\alpha$ = 3 at $T$ = $T_{BKT}$. Figure. 3(a) and 3(b) show the $I$-$V$ characteristics in the log-log scale of Ba$_{0.75}$ClNbS$_{2}$ and Ba$_{0.75}$ClNbSe$_{2}$ across the superconducting transition at zero field. For both compounds, the $I$-$V$ curve exhibits a crossover from a linear behavior corresponding to $\alpha$ = 1 in the normal state to a nonlinear behavior as the temperature decreases. By fitting the power law $V \propto I^{\alpha} (T)$ at various temperature, the BKT transition temperature $T_{BKT}$ = 0.99 K (1.2 K) is obtained where the exponent value $\alpha$ = 3 for Ba$_{0.75}$ClNbS$_{2}$ (Ba$_{0.75}$ClNbSe$_{2}$), as shown in Fig. 3(c) and 3(d). Moreover, in 2D superconductors, the temperature-dependent resistivity follows a formula $\rho(T) = \rho_{0}exp[-b/(T-T_{BKT})^{1/2}]$ when the temperature is close to the $T_{BKT}$, where $\rho_{0}$ and $b$ depend on the materials \cite{ref12,ref46}, which is another criterion for determining the $T_{BKT}$. The insets of Fig. 3(c) and 3(d) display the temperature dependence of resistivity for Ba$_{0.75}$ClNbS$_{2}$ and Ba$_{0.75}$ClNbSe$_{2}$ plotted in a $[dln(\rho)/dT]^{-2/3}$ scale. The extracted value of $T_{BKT}$ from the resistivity fitting is about 1.02 K (1.16 K) for Ba$_{0.75}$ClNbS$_{2}$ (Ba$_{0.75}$ClNbSe$_{2}$), consistent with the results of the $I$-$V$ analysis. Thereby, the intrinsic 2D superconductivity in Ba$_{0.75}$ClNbS$_{2}$ and Ba$_{0.75}$ClNbSe$_{2}$ has been verified by the observation of a BKT transition in these compounds, suggesting the introduction of Ba$_{0.75}$Cl insulation layer effectively weakens interlayer coupling and significantly changes electronic properties.

\begin{figure*}[htp]
    	\centering
    	\includegraphics[width=0.9\textwidth]{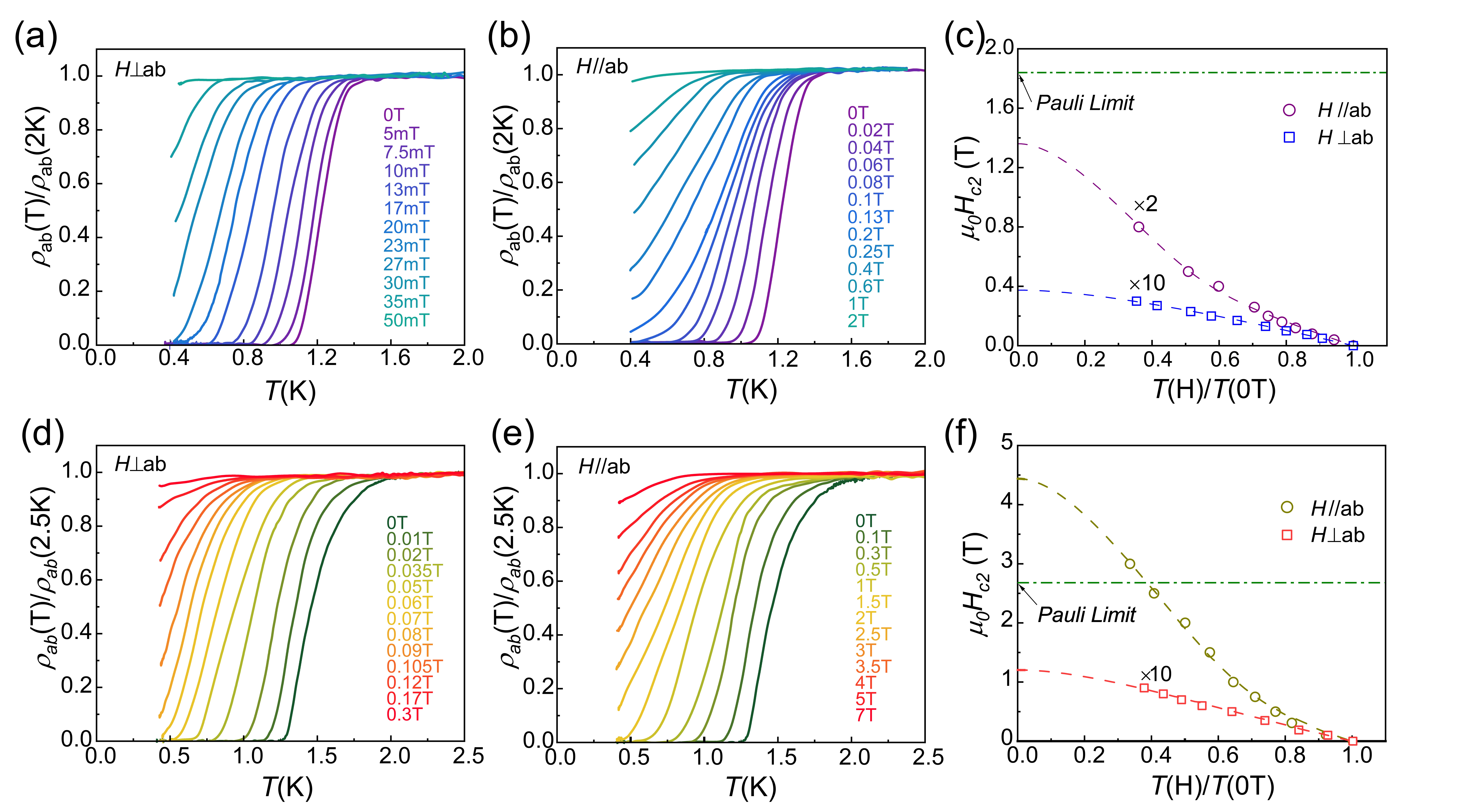}
    	\caption{Upper critical fields of Ba$_{0.75}$ClNbS$_{2}$ and Ba$_{0.75}$ClNbSe$_{2}$. (a) and (b) Resistivity $\rho(T)$ measured between 0.4 K and 2 K in Ba$_{0.75}$ClNbS$_{2}$ under magnetic fields applied perpendicular (a) and parallel (b) to the $ab$ plane. $\rho(T)$ is normalized to the value at 2 K. (c) Upper critical fields plotted against normalized temperature $T/T_{c}$ for Ba$_{0.75}$ClNbS$_{2}$. Blue and purple symbols represent $H_{c2}$ measured with $H$ perpendicular and parallel to the $ab$ plane, respectively; the out-of-plane and in-plane upper critical fields are multiplied by 10 and 2 for clarity, respectively. (d) and (e) Resistivity $\rho(T)$ measured between 0.4 K and 2.5 K in Ba$_{0.75}$ClNbSe$_{2}$ under magnetic fields applied perpendicular (d) and parallel (e) to the $ab$ plane. $\rho(T)$ is normalized to the value at 2.5 K. (f) Upper critical fields plotted against normalized temperature $T/T_{c}$ for Ba$_{0.75}$ClNbSe$_{2}$. Red and dark yellow symbols represent $H_{c2}$ measured with $H$ perpendicular and parallel to the $ab$ plane, respectively; the out-of-plane upper critical fields are multiplied by 10 for clarity. In (c) and (f), $H_{c2}$ is defined to be when $\rho(H)$ reaches 50\% of its normal state value; green dotted lines denote the Pauli paramagnetic limit. Dashed curves represent the result of fitting a two-band model of the upper critical field.
    	}
    	\label{FIG. 4}
\end{figure*}

	A large superconducting anisotropy $\gamma = H_{c2}^{\parallel ab}/H_{c2}^{\perp ab}$ is also a characteristic manifestation of the 2D nature of superconducting states, where $H_{c2}^{\perp ab}$ and $H_{c2}^{\parallel ab}$ are upper critical fields perpendicular and parallel to the $ab$-plane, respectively. We measured the temperature dependent resistivity under magnetic fields applied perpendicular [Fig. 4(a) and 4(d)] and parallel [Fig. 4(b) and 4(e)] to the $ab$ plane in Ba$_{0.75}$ClNbS$_{2}$ and Ba$_{0.75}$ClNbSe$_{2}$. The superconductivity is remarkably suppressed as the applied magnetic field increases. Indeed, a relatively low magnetic field can entirely obliterate the superconductivity at 0.4 K, when the magnetic field is oriented perpendicularly to the $ab$ plane. Figure. 4(c) and 4(f) show the normalized temperature ($T/T_{c}$) dependence of the upper critical field $H_{c2}(T)$ of Ba$_{0.75}$ClNbS$_{2}$ and Ba$_{0.75}$ClNbSe$_{2}$, defined as a field where the resistivity reaches 50\% of the normal state value. The relationships of $H$-$T$ display a positive curvature trend overall, which can be reasonable described by a two-band model: 
	\begin{align}
		ln(t)=&-[U_{1}(h)+U_{2}(h)+\frac{\lambda_{0}}{w}]/2\nonumber\\&+s[[U_{1}(h)-U_{2}(h)-\frac{\lambda_{-}}{w}]^{2}/4+\frac{\lambda_{12}\lambda_{21}}{w^{2}}]^{-1/2};\nonumber
\end{align}
	Where $t = T/T_{c}$, $U_{1,2}(h)=Re[\psi[1/2+(i+D_{1,2}/D_{0})h]]-\psi(1/2)$, $\psi(x)$ is the digamma function, $\lambda_{11}$ and $\lambda_{22}$ are the intraband BCS coupling constants, $\lambda_{12}$ and $\lambda_{21}$ are the interband BCS coupling constants, $\lambda_{\pm} = \lambda_{11} \pm \lambda_{22}$, $w = \lambda_{11}\lambda_{22} - \lambda_{12}\lambda_{21}$, $s = sign(w)$, $\lambda_{0} = (\lambda_{-}^{2} + 4\lambda_{12}\lambda_{21})^{1/2}$, $h = H_{c2}D_{1}/2\lambda_{0}T$, $D_{0} = \hbar/2m$, $\alpha = D_{0}/D_{1}$, $D_{1}$ and $D_{2}$ are the in-plane diffusivities of each band. The model takes both orbital and Zeeman pair breaking into account while considers negligible interband scattering \cite{ref47,ref48}. The parameters that best fits to the experimental data are summarized in Table S3. The upper critical fields at zero temperature of Ba$_{0.75}$ClNbS$_{2}$ are estimated to be $\mu_{0}H_{c2}^{\perp ab}(0) \cong$ 0.037 T and $\mu_{0}H_{c2}^{\parallel ab}(0) \cong$ 0.68 T, corresponding to $\gamma$ = 18.4. Intriguingly, a stronger anisotropy $\gamma$ = 37 is discovered in Ba$_{0.75}$ClNbSe$_{2}$, derived from the upper critical fields $\mu_{0}H_{c2}^{\perp ab}(0) \cong$ 0.12 T and $\mu_{0}H_{c2}^{\parallel ab}(0) \cong$ 4.44 T estimated from the model. These values of anisotropy are much higher than that of the parent 2$H$-NbS$_{2}$/NbSe$_{2}$, and comparable to that of other niobium dichalcogenides-based superlattices (Table S2) \cite{ref49,ref50}.

	For 2D superconductors with magnetic field applied in-plane, the orbital effect is strongly suppressed, and the Pauli paramagnetism becomes the predominant pair-breaking mechanism \cite{ref51,ref52}. Obviously, the $\mu_{0}H_{c2}^{\parallel ab}$ is below the value of Pauli limit in Ba$_{0.75}$ClNbS$_{2}$ [Fig. 4(c)]. However, the presence of finite momentum pairing or strong SOC can break the Pauli limit \cite{ref53}, resulting in an extraordinary in-plane upper critical field. Considering that there is stronger spin-orbit interaction in Ba$_{0.75}$ClNbSe$_{2}$ compared to Ba$_{0.75}$ClNbS$_{2}$, the $\mu_{0}H_{c2}^{\parallel ab}$ can be significantly enhanced due to possible potential Ising-type or Rashba-type SOC \cite{ref30,ref32}. As shown in Fig. 4(f), the Pauli limit is violated under the field applied parallel to the $ab$ plane (i.e., $\mu_{0}H_{c2}^{\parallel ab} > \mu_{0}H_{p}$, where the $\mu_{0}H_{p} \cong 1.84\times T_{c})$ in Ba$_{0.75}$ClNbSe$_{2}$. Moreover, the coherence lengths of both compounds are determined according to the anisotropy Ginzburg-Landau (GL) theory \cite{ref54,ref55}, following the relationship: $\mu_{0}H_{c2}^{\perp ab} = \phi_{0}/(2\pi\xi_{ab}^{2}), \mu_{0}H_{c2}^{\parallel ab} = \phi_{0}/(2\pi\xi_{ab}\xi_{\perp})$, where $\xi_{ab}$ and $\xi_{\perp}$are the in-plane and out-of-plane coherence lengths, respectively; $\phi_{0}$ is the magnetic flux quantum. The coherence lengths of Ba$_{0.75}$ClNbS$_{2}$ and Ba$_{0.75}$ClNbSe$_{2}$ are estimated: $\xi_{ab}$ $\approx$ 94.4 nm / 52.4 nm and $\xi_{\perp}$ $\approx$ 5.13 nm / 1.42 nm. In some cases, the $\xi_{\perp}$ could decrease to below the interlayer spacing d with the reduction of temperature in extremely anisotropic superconductors including the high-$T_{c}$ cuprates superconductor \cite{ref56}. The vortices are confined between the superconducting layers under in-plane fields, which possibly leads to a dimensional crossover characterized by a strong upturn curvature of $\mu_{0}H_{c2}^{\parallel ab} (T)$ as the temperature approaching $T_{c}$. Such features are also revealed in some TMD-based layered compounds with a large superconducting anisotropy \cite{ref32,ref57}. In particular, the $\xi_{\perp}$ is comparable to the distance (d $\approx$ 1.23 nm) between the superconducting layers in Ba$_{0.75}$ClNbSe$_{2}$, indicating the presence of a potential dimensional crossover, which further validates the intrinsic 2D superconductivity at low $T$.

	\section{Conclusion}	
	In summary, we have successfully synthesized two niobium dichalcogenides-based superconductors Ba$_{0.75}$ClNbS$_{2}$ and Ba$_{0.75}$ClNbSe$_{2}$. Through the comprehensive study of its transport measurement, the bulk superconductivity is observed with $T_{c}$ = 1 K and 1.25 K in both superlattices, respectively. The resistivity of normal state in comparison to the parent compounds exhibits a larger anisotropy due to the presence of the Ba$_{0.75}$Cl insulation layer. The intrinsic 2D superconductivity is revealed by a BKT transition and a remarkable superconducting anisotropy in Ba$_{0.75}$ClNbS$_{2}$ and Ba$_{0.75}$ClNbSe$_{2}$. Furthermore, the introduction of heavier Se atom in Ba$_{0.75}$ClNbSe$_{2}$ leads to a stronger spin-orbit interaction in superconducting state that breaks the Pauli limit. Our results significantly expand the path for the discovery of novel unconventional low-dimensional superconductors, and presenting an opportune moment to delve into a variety of low-dimensional quantum phases in bulk materials.

	\section*{Acknowledgements}
	We are grateful to Jianjun Ying, Zhenyu Wang and Tao Wu for fruitful discussions. We thank Zhongliang Zhu for experimental assistance. This work was supported by the National Key Research and Development Program of the Ministry of Science and Technology of China (Grants Nos. 2022YFA1602601 and 2022YFA1602602), the National Natural Science Foundation of China (Grants Nos. 12204448 and 12274390), the Innovation Program for Quantum Science and Technology (Grant No. 2021ZD0302802), the Anhui Initiative in Quantum Information Technologies (Grant No. AHY160000). M. Z. Shi. acknowledges the China Postdoctoral Science Foundation (2021M703107).


\begin{thebibliography}{99}
		\bibitem{ref1}  A. Shalnikov, Superconducting Thin Films, Nature \textbf{142}, 74 (1938).	
		\bibitem{ref2} D. B. Haviland, Y. Liu, and A. M. Goldman, Onset of superconductivity in the two-dimensional limit, Phys. Rev. Lett. \textbf{62}, 2180 (1989).
		\bibitem{ref3} M. Tinkham, Effect of fluxoid quantization on transitions of superconducting films, Phys. Rev. \textbf{129}, 2413 (1963).
		\bibitem{ref4} Y. Saito, T. Nojima, and Y. Iwasa, Highly crystalline 2D superconductors, Nat. Rev. Mater. \textbf{2}, 16094 (2017).
		\bibitem{ref5} Z. Li, L. Sang, P. Liu, Z. Yue, M. S. Fuhrer, Q. Xue, and X. Wang, Atomically Thin Superconductors, Small \textbf{17}, 1904788 (2021).
		\bibitem{ref6} C. Boix-Constant, S. Mañas-Valero, R. Córdoba, and E. Coronado, van der Waals heterostructures based on atomically thin superconductors, Adv. Electron. Mater. \textbf{7}, 2000987 (2021).
		\bibitem{ref7} Y. Xing, H. M. Zhang, H. L. Fu, H. W. Liu, Y. Sun, J. P. Peng, F. Wang, X. Lin, X. C. Ma, Q. K. Xue, J. Wang, and X. C. Xie, Quantum Griffiths singularity of superconductor-metal transition in Ga thin films, Science \textbf{350}, 542 (2015).
		\bibitem{ref8}  A. W. Tsen, B. Hunt, Y. D. Kim, Z. J. Yuan, S. Jia, R. J. Cava, J. Hone, P. Kim, C. R. Dean, and A. N. Pasupathy, Nature of the quantum metal in a two-dimensional crystalline superconductor, Nat. Phys. \textbf{12}, 208 (2016).
		\bibitem{ref9}  J. M. Kosterlitz, and D. J. Thouless, Ordering, metastability and phase transitions in two dimensional systems, J. Phys. C: Solid State Phys. \textbf{6}, 1181 (1973).
		\bibitem{ref10} A. F. Hebard, and A. T. Fiory, Evidence for the Kosterlitz-Thouless Transition in Thin Super-conducting Aluminum Films, Phys. Rev. Lett. \textbf{44}, 291 (1980).
		\bibitem{ref11} J. M. Lu, O. Zheliuk, I. Leermakers, N. F. Q. Yuan, U. Zeitler, K. T. Law, and J. T. Ye, Evidence for two-dimensional Ising superconductivity in gated MoS$_{2}$, Science \textbf{350}, 1353 (2015).
		\bibitem{ref12} N. Reyren, S. Thiel, A. D. Caviglia, L. F. Kourkoutis, G. Hammerl, C. Richter, C. W. Schneider, T. Kopp, A. S. Rüetschi, D. Jaccard, M. Gabay, D. A. Muller, J. M. Triscone, and J. Mannhart, Super-conducting Interfaces Between Insulating Oxides, Science \textbf{317}, 1196 (2007).
		\bibitem{ref13} S. Gariglio, M. Gabay, J. Mannhart, and J.-M. Triscone, Interface superconductivity, Physica C \textbf{514}, 189 (2015).
		\bibitem{ref14}  Y. Saito, Y. Kasahara, J. Ye, Y. Iwasa, and T. Nojima, Metallic ground state in an ion-gated two-dimensional superconductor, Science \textbf{350}, 409 (2015).
		\bibitem{ref15}  Y. Saito, Y. Nakamura, M. S. Bahramy, Y. Kohama, J. Ye, Y. Kasahara, Y. Nakagawa, M. Onga, M. Tokunaga, T. Nojima, Y. Yanase, and Y. Iwasa, Superconductivity protected by spin–valley locking in ion-gated MoS$_{2}$, Nat. Phys. \textbf{12}, 144 (2016). 
		\bibitem{ref16} D. Eom, S. Qin, M. Y. Chou, and C. K. Shih, Persistent Superconductivity in Ultrathin Pb Films: A Scanning Tunneling Spectroscopy Study, Phys. Rev. Lett. \textbf{96}, 027005 (2006).
		\bibitem{ref17} D. Jiang, T. Hu, L. X. You, Q. Li, A. Li, H. M. Wang, G. Mu, Z. Y. Chen, H. R. Zhang, G. H. Yu, J. Zhu, Q. J. Sun, C. T. Lin, H. Xiao, X. M. Xie, and M. H. Jiang, High-$T_{c}$ superconductivity in ultrathin Bi$_{2}$Sr$_{2}$CaCu$_{2}$O$_{8+x}$ down to half-unit-cell thickness by protection with graphene, Nat. Commun. \textbf{5}, 5708 (2014).
		\bibitem{ref18}  S. C. de la Barrera, M. R. Sinko, D. P. Gopalan, N. Sivadas, K. L. Seyler, K. Watanabe, T. Taniguchi, A. W. Tsen, X. Xu, D. Xiao, and B. M. Hunt, Tuning Ising superconductivity with layer and spin–orbit coupling in two-dimensional transition metal dichalcogenides, Nat. Commun. \textbf{9}, 1427 (2018).
		\bibitem{ref19}  A. Devarakonda, H. Inoue, S. Fang, C. Ozsoy-Keskinbora, T. Suzuki, M. Kriener, L. Fu, E. Kaxiras, D. C. Bell, and J. G. Checkelsky, Clean 2D superconductivity in a bulk van der Waals superlattice, Science \textbf{370}, 231 (2020).
		\bibitem{ref20} A. Devarakonda, A. Chen, S. Fang, D. Graf, M. Kriener, A. J. Akey, D. C. Bell, T. Suzuki and J. G. Checkelsky, Evidence of striped electronic phases in a structurally modulated superlattice, Nature \textbf{631}, 526 (2024).
		\bibitem{ref21} G. Li, A. Luican, J. M. B. Lopes dos Santos, A. H. Castro Neto, A. Reina, J. Kong, and E. Y. Andrei, Observation of Van Hove singularities in twisted graphene layers, Nat. Phys. \textbf{6}, 109 (2010).
		\bibitem{ref22}M. Yankowitz, S. Chen, H. Polshyn, Y. Zhang, K. Watanabe, T. Taniguchi, D. Graf, A. F. Young, and C. R. Dean, Tuning superconductivity in twisted bilayer graphene, Science \textbf{363}, 1059 (2019).
		\bibitem{ref23} T. S. Khaire, M. A. Khasawneh, W. P. Pratt, and N. O. Birge, Observation of spin-triplet super-conductivity in Co-based Josephson junctions, Phys. Rev. Lett. \textbf{104}, 137002 (2010).
        \bibitem{ref24} L. Balents, C. R. Dean, D. K. Efetov, and A. F. Young, Superconductivity and strong correlations in moiré flat bands, Nat. Phys. \textbf{16}, 725 (2020).
        \bibitem{ref25} C. Jin, E. C. Regan, A. Yan, M. I. B. Utama, D. Wang, S. Zhao, Y. Qin, S. Yang, Z. Zheng, S. Shi, K. Watanabe, T. Taniguchi, S. Tongay, A. Zettl, and F. Wang, Observation of moiré excitons in WSe$_{2}$/WS$_{2}$ heterostructure superlattices, Nature \textbf{567}, 76 (2019).
         \bibitem{ref26} B. Zhao, D. Shen, Z. Zhang, P. Lu, M. Hossain, J. Li, B. Li, and X. Duan, 2D Metallic Transition–Metal Dichalcogenides: Structures, Synthesis, Properties, and Applications, Adv. Funct. Mater. \textbf{31}, 2105132 (2021).
        \bibitem{ref27} H. T. Wang, H. T. Yuan, S. S. Hong, Y. B. Li, and Y. Cui, Physical and chemical tuning of two-dimensional transition metal dichalcogenides, Chem. Soc. Rev. \textbf{44}, 2664 (2015).
		\bibitem{ref28} L. Du, T. Hasan, A. Castellanos-Gomez, G.-B. Liu, Y. Yao, C. Ning Lau, and Z. Sun, Engineering symmetry breaking in 2D layered materials, Nat. Rev. Phys. \textbf{3}, 193 (2021).
        \bibitem{ref29} H. X. Zhang, A. Rousuli, K. Zhang, L. Luo, C. Guo, X. Cong, Z. Lin, C. Bao, H. Zhang, S. Xu, R. Feng, S. Shen, K. Zhao, W. Yao, Y. Wu, S. Ji, X. Chen, P. Tan, Q.-K. Xue, Y. Xu, W. Duan, P. Yu, and S. Zhou, Tailored Ising superconductivity in intercalated bulk NbSe$_{2}$, Nat. Phys. \textbf{18}, 1425 (2022).
		\bibitem{ref30} X. Xi, Z. Wang, W. Zhao, J.-H. Park, K. T. Law, H. Berger, L. Forró, J. Shan, and K. F. Mak, Ising pairing in superconducting NbSe$_{2}$ atomic layers, Nat. Phys. \textbf{12}, 139 (2016).
		\bibitem{ref31} R. Sun, J. Deng, X. Wu, M. Hao, K. Ma, Y. Ma, C. Zhao, D. Meng, X. Ji, and Y. Ding, High anisotropy in electrical and thermal conductivity through the design of aerogel-like superlattice (NaOH)$_{0.5}$NbSe$_{2}$, Nat. Commun. \textbf{14}, 6689 (2023).
		\bibitem{ref32} P. Samuely, P. Szabó, J. Kaˇcmarˇcík, A. Meerschaut, L. Cario, A. G. M. Jansen, T. Cren, M. Kuzmiak, O. Šofranko, and T. Samuely, Extreme in-plane upper critical magnetic fields of heavily doped quasi-two-dimensional transition metal dichalcogenides, Phys. Rev. B \textbf{104}, 224507 (2021).
		\bibitem{ref33} X. Yang, T. Yu, C. Xu, J. Wang, W. Hu, Z. Xu, T. Wang, C. Zhang, Z. Ren, Z. A. Xu, M. Hirayama, R. Arita, and X. Lin, Anisotropic superconductivity in topological crystalline metal Pb$_{1/3}$TaS$_{2}$ with multiple Dirac fermions, Phys. Rev. B \textbf{104}, 035157 (2021).
		\bibitem{ref34}  CrysAlisPro Software system. Version 1.171.37.35, (Agilent Technologies Ltd, Yarnton, Oxfordshire, England, 2014).
        \bibitem{ref35} L. J. Bourhis, O. V. Dolomanov, R. J. Gildea, J. A. K. Howard, and H. Puschmann, Olex2: a complete structure solution, refinement and analysis program, J. Appl. Crystallogr. \textbf{42}, 339 (2009).
        \bibitem{ref36} G. M. Sheldrick, SHELX--Integrated space-group and crystal-structure determination, Acta Crystallogr. Sect. A \textbf{71}, 3 (2015).
        \bibitem{ref37} G. M. Sheldrick, A short history of SHELX, Acta Crystallogr. A \textbf{64}, 112 (2008).
        \bibitem{ref38} See Supplemental Material at XX for a detailed structural information, the results of resistivity anisotropy, magnetoresistance and Hall measurements and the fitting parameter of upper critical field, which includes Refs. [19,39-43,49,50].		
        \bibitem{ref39} K. Onabe, M. Naito, and S. Tanaka, Anisotropy of Upper Critical Field in Superconducting 2H-NbS$_{2}$, J. Phys. Soc. Jpn. \textbf{45}, 50 (1978).
		\bibitem{ref40} M. Naito and S. Tanaka, Electrical Transport Properties in 2H-NbS$_{2}$, -NbSe$_{2}$, -TaS$_{2}$ and -TaSe$_{2}$, J. Phys. Soc. Jpn. \textbf{51}, 219 (1982).
		\bibitem{ref41} R. Yan, G. Khalsa, B. T. Schaefer, A. Jarjour, S. Rouvimov, K. C. Nowack, H. G. Xing, and D. Jena, Thickness dependence of superconductivity in ultrathin NbS$_{2}$, Appl. Phys. Express \textbf{12}, 023008 (2019).
		\bibitem{ref42} B. W. Pfalzgraf and H. Speckels, The anisotropy of the upper critical field $H_{c2}$ and electrical resistivity in 2H-NbS$_{2}$, J. Phys. C \textbf{20}, 4359 (1987).
		\bibitem{ref43} A. LeBlanc and A. Nader, Resistivity anisotropy and charge density wave in 2H-NbSe$_{2}$ and 2H-TaSe$_{2}$, Solid State Commun. \textbf{150}, 1346 (2010).
		\bibitem{ref44} P. Minnhagen, Kosterlitz-Thouless transition for a two-dimensional superconductor: Magnetic-field dependence from a Coulomb-gas analogy, Phys. Rev. B \textbf{23}, 5745 (1981).
		\bibitem{ref45} P. Minnhagen, The two-dimensional coulomb gas, vortex unbinding, and superfluid super-conducting films, Rev. Mod. Phys. \textbf{59}, 1001 (1987).
		\bibitem{ref46} B. I. Halperin and D. R. Nelson, Resistive transition in superconducting films, J. Low Temp. Phys. \textbf{36}, 599 (1979).
		\bibitem{ref47}  A. Gurevich, Enhancement of the upper critical field by nonmagnetic impurities in dirty two-gap superconductors, Phys. Rev. B \textbf{67}, 184515 (2003).
		\bibitem{ref48} J. Jaroszynski, F. Hunte, L. Balicas, Youn-jung Jo, I. Raičević, A. Gurevich, D. C. Larbalestier, F. F. Balakirev, L. Fang, P. Cheng, Y. Jia, and H. H. Wen, Upper critical fields and thermally-activated transport of NdFeAsO$_{0.7}$F$_{0.3}$ single crystal, Phys. Rev. B \textbf{78}, 174523 (2008).
		\bibitem{ref49} F. Soto, H. Berger, L. Cabo, C. Carballeira, J. Mosqueira, D. Pavuna, P. Toimil, and F. Vidal, Electric and magnetic characterization of NbSe$_{2}$ single crystals: Anisotropic superconducting fluctuations above $T_{C}$, Phys. C \textbf{460-462}, 789 (2007).
		\bibitem{ref50}  K. Ma, S. Jin, F. Meng, Q. Zhang, R. Sun, J. Deng, L. Chen, L. Gu, G. Li, and Z. Zhang, Two-dimensional superconductivity in a bulk superlattice van der Waals material Ba$_{6}$Nb$_{11}$Se$_{28}$, Phys. Rev. Mater. \textbf{6}, 044806 (2022).
		\bibitem{ref51} A. M. Clogston, Upper limit for critical field in hard superconductors, Phys. Rev. Lett. \textbf{9}, 266 (1962).
		\bibitem{ref52} B. S. Chandrasekhar, A note on the maximum critical field of high-field superconductors, Appl. Phys. Lett. \textbf{1}, 7 (1962).
		\bibitem{ref53} Y. Cao, J. M. Park, K. Watanabe, T. Taniguchi, and P. Jarillo Herrero, Pauli-limit violation and re-entrant superconductivity in moiré graphene, Nature \textbf{595}, 526 (2021).
		\bibitem{ref54} M. Tinkham, Introduction to Superconductivity (Dover Publications, 2004).
		\bibitem{ref55} V. L. Ginzburg and L. D. Landau, On Superconductivity and Superfluidity: A Scientific Autobiography (Springer, Berlin, 2009).
		\bibitem{ref56} T. Schneider and A. Schmidt, Dimensional crossover in the upper critical field of layered superconductors, Phys. Rev. B \textbf{47}, 5915 (1993).
		\bibitem{ref57} R. V. Coleman, G. K. Eiserman, S. J. Hillenius, A. T. Mitchell, and J. L. Vicent, Dimensional crossover in the superconducting intercalated layer compound 2H-TaS$_{2}$, Phys. Rev. B \textbf{27}, 125 (1983).
		
	\end{thebibliography}
\end{document}